# The Effect of Age Introduced to Virtual Reality on Susceptibility to Motion Sickness

Michael Abia

## Abstract

Every human with a functioning vestibular system is capable of feeling motion sickness, but some are more vulnerable than others. Based on the leading theories explaining this condition, vulnerability should be predicted by a person's years of real-life experience before using a VR device and years of VR experience after. A questionnaire was filled out on susceptibility to motion sickness in VR by people on VR-related forums. Results from the survey show that the condition has a significant relationship with age or experience outside the environment.

Keywords: Motion Sickness, Virtual Reality, Age, Experience

## Introduction

Motion Sickness is a condition that occurs due to motion or the perception of it, and affects people with symptoms like dizziness, nausea and discomfort. Known as car-sickness to drivers, sea-sickness to sailors, air-sickness to pilots, space-sickness to astronauts and simulator sickness to virtual reality gamers. The leading theories explain how or why it occurs, without necessarily contradicting each other.
Sensory Conflict Theory proposes that motion sickness is caused by a mismatch between actual and perceived motion[1]. Critics of this theory proposed the Postural Instability Theory, suggesting that sensory conflict is an unreliable model for producing motion sickness symptoms and that postural imbalance is more measurable[2,3]. The sensory conflict theory is also guilty of only explaining motion sickness in people who suffer the condition, hence why a variant of this theory was introduced. The Neural Mismatch Theory explains that motion sickness is caused by unfamiliar sensory patterns, so people suffer from motion sickness if they are experiencing an unusual sequence of physical or perceived movements[4].

The Neural Mismatch Theory proposes that contradiction with our previous experiences is the cause of our intolerance to motion. It seems a reasonable deduction in line with other theories given that postural instability is expected with inexperience in an environment or condition, and even when subject to the same sensory conflicts some people are less vulnerable than others. It is widely agreed that over time and exposure, we adapt to new environments, activities or even substances that might have been previously uncomfortable[5,6]. However, it is not clear whether age (or experience) before and with VR plays a role in the person's susceptibility to motion sickness.

**Method**

We designed the survey to collect data including age when introduced to virtual reality, their current subjective rate of motion sickness (rarely/frequently), and years of experience using VR devices. All participants were required to have used a VR device and all questions were required to be answered before submitting the form. The questionnaire was posted on the r/virtualreality, r/vrchat, r/vrdev, r/learnvrdev, r/oculus, r/valveindex, and r/psvr sections of Reddit.com from the 1st of April 2022 to the 30th of April 2022. Data was collected through Google Forms and saved to Google Sheets where it was analyzed. For our analysis, we used the student t-test separately for both the mean age introduced to VR and the mean years of experience in VR.

**Result**

173 people filled the survey, their mean age introduced to virtual reality was 27.99 years (SD = 11.49) and mean years of experience was 2.60 years (SD = 2.26). 28.9% described themselves as frequently motion sick with their mean age introduced at 32.22 (SD = 12.75) and mean years of experience since then at 2.28 (SD = 2.19). 71.1% claim they are rarely motion sick with the mean age introduced at 26.28 (SD = 10.51) and the mean years of experience since then at 2.73 (SD = 2.28).

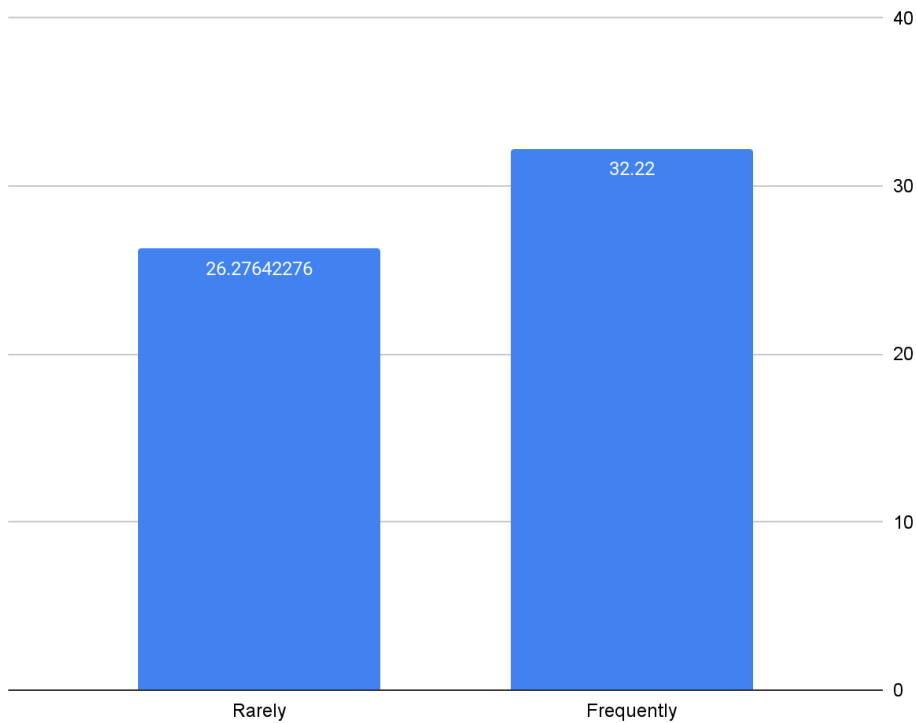

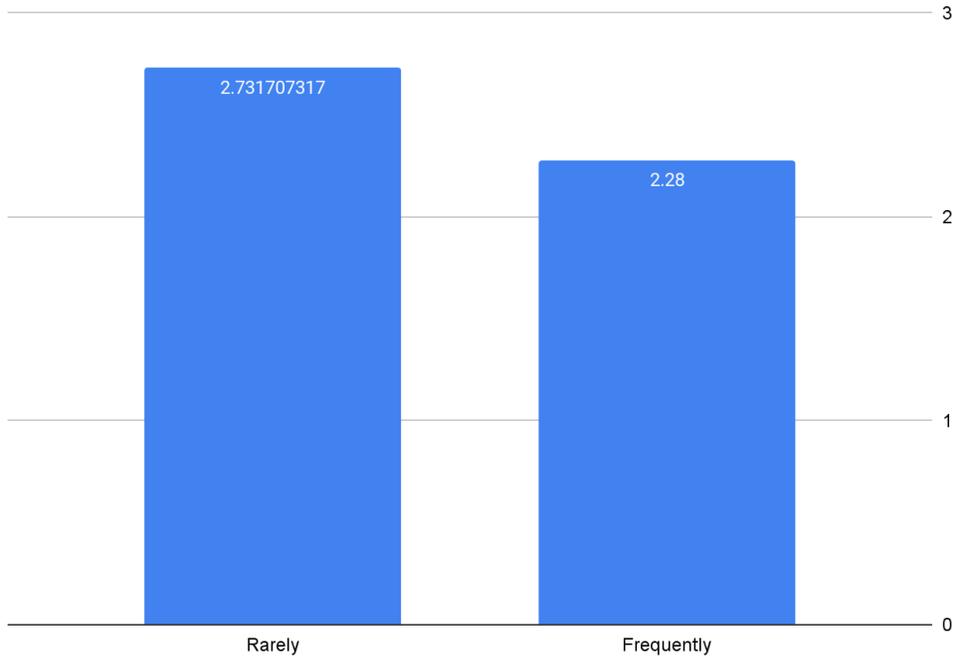

| Summary Statistics | n | Mean | Median | Mode | Standard Deviation | Variance | p-value |
|---|---|---|---|---|---|---|---|
| All (Age Introduced) | 173 | 27.99 | 26 | 21 | 11.49 | 131.97 | |
| Frequently | 50 | 32.22 | 31 | 40 | 12.75 | 162.58 | 0.0046 |
| Rarely | 123 | 26.28 | 25 | 21 | 10.51 | 110.46 | |
| All (Experience) | 173 | 2.60 | 2 | 1 | 2.26 | 5.10 | |
| Frequently | 50 | 2.28 | 1.5 | 0 | 2.19 | 4.82 | 0.2281 |
| Rarely | 123 | 2.73 | 2 | 1 | 2.28 | 5.20 | |

From statistical tests we found the ages introduced to VR for the frequently motion sick vs the rarely motion sick to be significantly different (p = 0.0046). But the variation in the years of experience using VR between the frequently motion sick and the rarely motion sick was high, setting a low chance of a significant relationship (p = 0.2281).

## Discussion

Among active VR users, we have found that people who tend to feel that they are not prone to motion sickness are younger than people who feel that they are. However we could not find the difference in years of VR experience between the frequently and rarely motion sick reliable. Past research may have found significant relationships between age and motion sickness, but this study divides age pre-environment and post-environment to remove the effect of habituation on their susceptibility to motion sickness. It is possible that other discriminatory factors are responsible for these results. Like the distance between the headset lens(IPDs), which was found to be the main cause of difference in motion sickness vulnerability between males and females[7]. It is also worthy to note the possibility that those with a lot of VR experience who frequently get motion sickness might have conditioned themselves into it without realizing their IPD setting was wrong[8].

Several variables that might influence motion sickness were not assessed such as individual physical or biological differences, variations in these experiences, learning abilities were not considered. Future research should investigate the effect of a more specific experience on susceptibility to motion sickness in that controlled environment.

## Conclusion

We have observed that within the active VR communities found on Reddit, people who consider themselves vulnerable to motion sickness were significantly more experienced (age) before using VR than those who consider themselves resistant. Having lived more years without using VR seems to predict that an individual will likely be prone to motion sickness in VR. Confirming the Neural Mismatch theory's proposition that contradiction to previous experiences leads to motion sickness.

## Additional Documents

Survey Questions:
https://docs.google.com/forms/d/e/1FAIpQLScGw6roMZgDXqGnAhPxnBuxoQBWIfnF6kothI-eM3EiCSr9_g/viewform?usp=sf_link

Survey Posts:
- [Motion Sickness Research : r/virtualreality](#)
- [Please Help! - Motion Sickness Research : r/VRchat](#)
- [Motion Sickness Research : r/vrdev](#)
- [Motion Sickness Research : r/learnVRdev](#)
- [Motion Sickness Research : r/oculus](#)
- [Motion Sickness Research : r/ValveIndex](#)
- [Motion Sickness Research : r/PSVR](#)

Survey Responses:
📊 Motion Sickness Survey (Responses)
📊 Motion Sickness Summary Statistics

## References

1. Claremont, C. A., (1931). The psychology of seasickness. *Psyche*, *11*, 86-90.
2. Stoffregen, T. A., & Riccio, G. E. (1991). An Ecological Critique of the Sensory Conflict Theory of Motion Sickness. *Ecological Psychology*, *3(3)*, 159-194.
3. Bonnet, C. T., Faugloire, E., Riley, M. A., Bardy, B. G., & Stoffregen, T. A. (2006). Motion sickness preceded by unstable displacements of the center of pressure. *Human movement science*, *25(6)*, 800–820.
4. Reason, J. T., (1978). Motion sickness adaptation: a neural mismatch model. *Journal of the Royal Society of Medicine*, *71(11)*, 819–829.
5. Broom, D. M., (2006). Adaptation. *Berl Munch Tierarztl Wochenschr*, *119(1-2)*, 1–6.
6. Lafontaine, M. P., Knoth, I. S., & Lippe, S., (2020). Learning Abilities. *Handbook of clinical neurology*, *173*, 241–254.
7. Stanney, K., Fidiopiastis, C., & Foster, L., (2019). Virtual Reality is Sexist: But It Does Not Have To Be. *Front. Robot. AI*, *7*, 4.
8. Toschi, N., Kim, J., Sclocco, R., Duggento, A., Barbieri, R., Kuo, B., & Napadow, V., (2017). Motion sickness increases functional connectivity between visual motion and nausea-associated brain regions. *Autonomic Neuroscience*, *202*, 108-133.